# Automated identification and quantification of myocardial inflammatory infiltration in digital histological images to diagnose myocarditis

**Authors**
Yanyun Liu[a,b,c#]
yanyunliu@stu.xidian.edu.cn
Xiumeng Hua[d,e,f,g#]
huaxiumeng@163.com
Congrui Wang[d,e,f,g]
crwang99@163.com
Xiao Chen[d,e,f,g]
584664708@qq.com
Yu Shi[a,b,c]
shiyu_95@stu.xidian.edu.cn
Shouping Zhu[a,b,c]
spzhu@xidian.edu.cn
Jiangping Song[d,e,f,g*]
fwsongjiangping@126.com
Weihua Zhou[h,i]
whzhou@mtu.edu

**Institutions**
a. School of Life Science and Technology, Xidian University & Engineering Research Center of Molecular and Neuro Imaging, Ministry of Education, Xi'an, Shaanxi, China
b. Xi'an Key Laboratory of Intelligent Sensing and Regulation of Trans-Scale Life Information & International Joint Research Center for Advanced Medical Imaging and Intelligent Diagnosis and Treatment, School of Life Science and Technology, Xidian University, Xi'an, Shaanxi, China
c. Innovation Center for Advanced Medical Imaging and Intelligent Medicine, Guangzhou Institute of Technology, Xidian University, Guangzhou, Guangdong, China
d. Beijing Key Laboratory of Preclinical Research and Evaluation for Cardiovascular Implant Materials, Animal Experimental Centre, Fuwai Hospital, National Centre for Cardiovascular Disease, Chinese Academy of Medical Sciences and Peking Union Medical College, Beijing, China.
e. State Key Laboratory of Cardiovascular Disease, Fuwai Hospital, National Center for Cardiovascular Diseases, Chinese Academy of Medical Sciences and Peking Union Medical College, 167A Beilishi Road, Xi Cheng District, Beijing, China.
f. Department of Cardiovascular Surgery, Fuwai Hospital, National Center for Cardiovascular Diseases, National Clinical Research Center of Cardiovascular Diseases, State Key Laboratory of Cardiovascular Disease, Chinese Academy of Medical Sciences and Peking Union Medical College, Beijing, China.
g. The Cardiomyopathy Research Group, Fuwai Hospital, Beijing, China.

h.  Department of Applied Computing, Michigan Technological University, Houghton, MI, USA
i.  Center for Biocomputing and Digital Health, Institute of Computing and Cybersystems, and Health Research Institute, Michigan Technological University, Houghton, MI, USA

#Yanyun Liu and Xiumeng Hua contributed equally to this work. Yanyun Liu: conception design, analysis, and interpretation of data; drafting of the manuscript and revising it critically for important intellectual content. XiuMeng Hua: active involvement in collecting data and performing experiments with subsequent participation in data analysis, drafting the manuscript, and revising it critically for important intellectual content.

**\*Address for correspondence**
Shouping Zhu
Tel.: +86-029-81891070
E-mail: spzhu@xidian.edu.cn
Mailing address: School of Life Science and Technology, Xidian University, Xi'an, Shaanxi 710126, China.

Or
Jingping Song
Tel.:
E-mail address: fwsongjiangping@126.com
Mailing address: State Key Laboratory of Cardiovascular Disease, Fuwai Hospital, National Center for Cardiovascular Diseases, Chinese Academy of Medical Sciences and Peking Union Medical College, 167A Beilishi Road, Xi Cheng District, Beijing 100037, China.

# Abstract

**Background**

Given the dimensionality of WSIs and the increase in the number of potential cases, the manual histological diagnosis of myocarditis based on Dallas Criteria is time-consuming, experience-dependent.

**Objectives**

To overcome the limitations of manual diagnosis methods, this study aims to develop a new computational pathology approach that automates the identification and quantification of myocardial inflammatory infiltration in digital hematoxylin and eosin (HE)-stained images to assist pathologists in diagnosing myocarditis.

**Methods**

898 HE-stained whole slide images (WSIs) of myocardium from 154 heart transplant patients diagnosed with myocarditis or dilated cardiomyopathy (DCM) were included in this study. An automated deep-learning (DL)-based computational pathology approach was developed to identify nuclei and detect myocardial inflammatory infiltration, enabling the quantification of the lymphocyte nuclear density (LND) on myocardial WSIs. A cutoff value based on the quantification of LND was proposed to determine if the myocardial inflammatory infiltration was present. The performance of our approach was evaluated with a five-fold cross-validation experiment, tested with an internal test set from the myocarditis group, and confirmed by an external test from a double-blind trial group.

**Results**

An LND of $1.02/mm^2$ could distinguish WSIs with myocarditis from those without. The accuracy, sensitivity, specificity, and area under the receiver operating characteristic curve (AUC) in the five-fold cross-validation experiment were $0.899 \pm 0.035$, $0.971 \pm 0.017$, $0.728 \pm 0.073$ and $0.849 \pm 0.044$, respectively. For the internal test set, the accuracy, sensitivity, specificity, and AUC were 0.887, 0.971, 0.737, and 0.854, respectively. The accuracy, sensitivity, specificity, and AUC for the external test set reached 0.853, 0.846, 0.858, and 0.852, respectively.

**Conclusion**

Our new approach provides accurate and reliable quantification of the LND of myocardial WSIs based on myocardial inflammatory infiltration, facilitating automated quantitative diagnosis of myocarditis with HE-stained images.

# 1. Introduction

The gold standard for diagnosing myocarditis is a histopathological evaluation that meets the "Dallas Criteria"[1–4]. The Dallas criteria require an inflammatory infiltrate associated with myocyte necrosis or damage not characteristic of an ischemic event[3–6]. However, this method is limited by the high interobserver variability in interpreting biopsy specimens[3]. Therefore, developing an objective approach to quantifying myocardial inflammatory infiltration in the widely used hematoxylin and eosin (HE)-stained images is essential.

In HE-stained images, lymphocyte nuclear aggregation is the primary indicator of myocardial inflammatory infiltration. Therefore, quantifying lymphocyte nuclear density (LND) may be a potential predictor for the diagnosis of myocarditis. The routine pathological diagnosis of myocarditis using HE-stained images currently depends on manual examination by pathologists, making it impractical to perform manual quantification of LND on high-resolution gigapixel images. Digital whole slide imaging (WSI) technology has revolutionized the storage and analysis of pathological images[7]. With the development of machine learning (ML) and deep learning (DL), computational pathology introduces a new way for the quantitative analysis of digital whole slide images (WSIs)[8–11], improving the accuracy and efficiency of pathological diagnosis significantly[9]. It also helps minimize inter-observer variability, promoting a more standardized diagnostic process. However, no DL-based computational pathology approach has been proposed to identify and quantify myocardial inflammatory infiltration in HE-stained images.

In this paper, we aimed to develop a new DL-based computational pathology approach to automatically identify nuclei and myocardial inflammatory infiltration, allowing quantification of LND in HE-stained images.

## 2. Materials and Methods

### 2.1 Study cohort and design

The cohort comprised histological slice data from 154 heart transplant patients treated at the Fuwai Hospital, Chinese Academy of Medical Sciences, and Peking Union Medical College between 2015 and 2022. This study was approved by the ethics committee of The Fuwai Hospital, Chinese Academy of Medical Sciences, and Peking Union Medical College. These patients were diagnosed with either myocarditis or dilated cardiomyopathy (DCM). Each patient contributed one or more slices from the left ventricle, right ventricle, and interventricular septum, resulting in 910 digital pathological images. The standard section procedure was subsequently employed for HE-staining. Digital WSIs were captured using a Zeiss microscope at 40x magnification, with a pixel size of 0.22μm. WSIs with uneven staining and blurry features were excluded through a quality control process, resulting in 898 WSIs available for further analysis. The dataset consisted of the myocarditis group and the double-blind trial group.

In the myocarditis group, 44 individuals with a total of 253 WSIs were randomly assigned to the training set (80%, n = 35, 200 WSIs with 47 negatives and 153 positives, where "n" represents the patient number) and the internal test set (20%, n = 9, 53 WSIs with 15 negatives and 38 positives) for algorithm development. The double-blind trial group, which included 110 patients (645 WSIs with 353 negatives and 292 positives) diagnosed with myocarditis or DCM, served as the external test group. These 110 patients were continuously enrolled from Fuwai hospital Heart Transplantation Center with pathological diagnosis by three clinical pathologist. Pathology diagnosis criterion was followed by the guideline from ESC/AHA[12,13]. **Figure 1** provides an overview of the distribution and utilization of the study data.

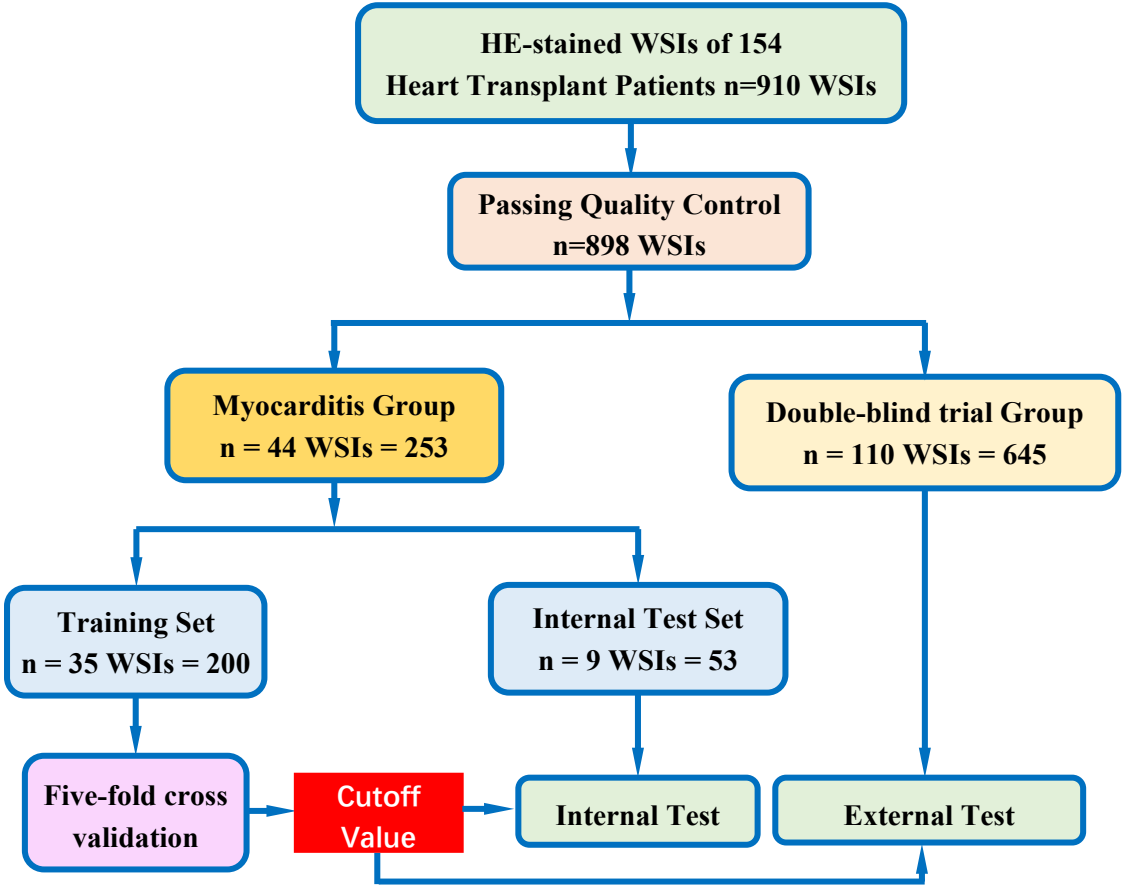

**Figure 1.** A summary of the flowchart illustrates the cohort distribution, with the letter 'n' representing the number of patients.

## 2.2 Measurement of LND

WSIs typically have incredibly high resolution, often in the gigapixel range, posing challenges for direct analysis due to memory limitations. Therefore, it is a common practice[14] to divide WSIs into smaller patches for analysis, as illustrated in **Figure 2**. To enhance the computational efficiency and reduce storage memory requirements, the WSIs in our study were down-sampled by a factor of 64. Low-resolution images were converted from RGB space to HSV color space. Median filtering was applied to the HSV image to separate the myocardial tissue from the background. This filtering process helped reduce noise and improve the accuracy of subsequent segmentation. Then, the Otsu algorithm[15] was employed for image thresholding, automatically determining an optimal threshold to separate the myocardial tissue from the rest of the image. The resulting segmented tissue regions were utilized to obtain the coordinates of patches for further analysis. These patches were then converted back to the original high-resolution image, allowing for the extraction of patches with the size of 1,024 x 1,024 that contained relevant tissue information. This step ensured that the analysis focused on specific regions of myocardial tissue within the WSIs while preserving the necessary details for quantification.

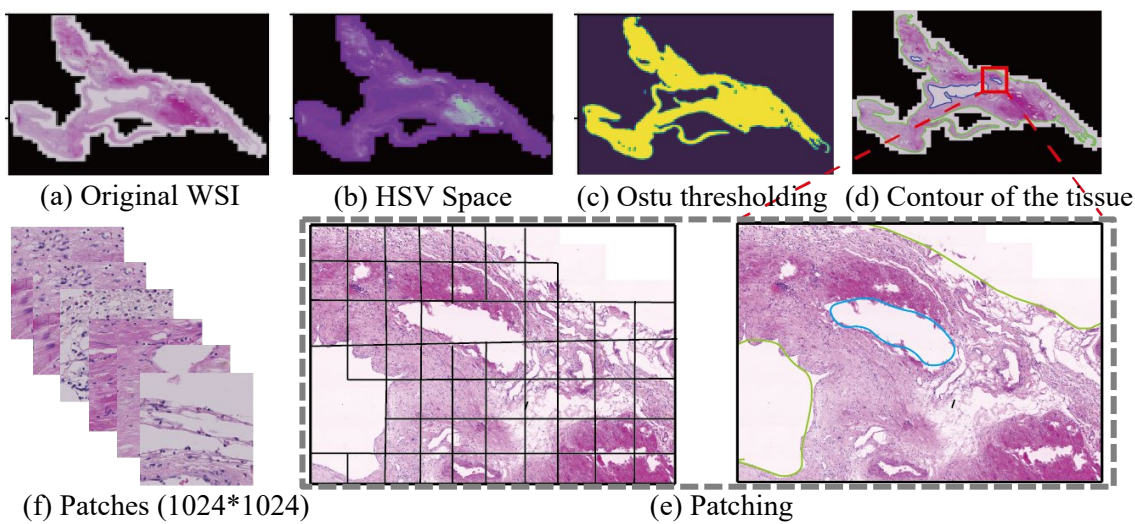

(a) Original WSI  (b) HSV Space  (c) Ostu thresholding  (d) Contour of the tissue

(f) Patches (1024*1024)  (e) Patching

**Figure 2**. The preprocessing procedure of WSIs. (a) and (b) are the low-resolution WSIs in RGB and HSV color spaces, respectively, (c) shows the results of Otsu thresholding, (d) shows the contour of the myocardial tissue, (e) provides a closer look of (d), (f) shows the final patches

Nuclei identification within each patch was performed using StarDist[16], a lightweight DL network. StarDist combines DL and a star-convex polygon representation to achieve precise and reliable nuclei segmentation in oncology. To assess its performance in the myocardial field, ten representative patches were selected from a diverse range of patients to evaluate the nuclei detection accuracy. These patches covered different myocarditis types and various locations within the heart. Three expert pathologists conducted a thorough examination and manually annotated the results as the reference standard, which was used to evaluate the accuracy of nuclei identification.

Based on the results of nuclei identification, we identified the inflammatory infiltration[17]. Firstly, a thresholding method based on the radius was employed to exclude some nuclei with a large radius, which may be myocardial nuclei or fiber nuclei. Next, the Euclidean distance between the remaining nuclei was calculated, and the proximity diagram between the nuclei was generated. When the number of aggregated nuclei in a proximity diagram is greater than or equal to 14[18], these nuclei are the lymphocyte nuclei and this diagram is identified as an inflammatory infiltration. To evaluate the accuracy of this algorithm in identifying inflammatory infiltration, we randomly selected ten WSIs from ten patients. Within these WSIs, an experienced pathologist manually identified patches containing inflammatory infiltration as the ground truth. The accuracy of our automatic identification algorithm was thus evaluated.

Finally, the LND was obtained by calculating the average number of lymphocyte nuclei per square millimeter of myocardial tissue on WSIs.

## 2.3 Diagnosis of myocarditis with LND

To investigate whether LND could be used as a predictor for myocarditis diagnosis, we evaluated the performance of different LND values for diagnosing myocarditis. A five-fold cross-validation was conducted to test the robustness of the optimal cutoff value for LND-based diagnosis of myocarditis. In the training set, the cohort of 35 patients with 200 WSIs was randomly divided into five subsets. Each subset consisted of 27 to 32 WSIs from seven patients. The internal and external test sets were used to further assess the feasibility of using LND to diagnose myocarditis at the WSI and patient levels.

The accuracy, sensitivity, specificity, and area under the receiver operating characteristic (ROC) curve (AUC), were employed to evaluate the performance of LND as a diagnostic predictor for myocarditis.

## 3. Results

**3.1 Study population**

The baseline characteristics of the 44 patients in the myocarditis group were shown in **Table 2**. The gender distribution was approximately equal. The mean ages of onset and transplantation were 36.54 and 42.75 years, respectively. As reported, viral infection was the cause of myocarditis, and 25% of patients were diagnosed with viral myocarditis. The prevalence of comorbidities such as hypertension was relatively low. Furthermore, 77.3% of these patients had NYHA class III or IV cardiac function. These patients were characteristic with the enlarged ventricle (LVEDD: $63.60 \pm 14.52$) and low cardiac function ($32.00 \pm 13.84$).

Table 2 The baseline characteristics of the myocarditis group

| Characteristics | Myocarditis Group |
|---|---|
| Male | 25 (57%) |
| Age, years | $43.75 \pm 14.99$ |
| Age of onset, years | $36.54 \pm 14.69$ |
| Age of transplantation, years | $42.75 \pm 15.10$ |
| Viral infection (n, %) | 11 (25%) |
| Myocardial infarction (n, %) | 3 (6.8%) |
| Hypertension (n, %) | 3 (6.8%) |
| Hyperthyroidism (n, %) | 1 (2.3%) |
| Myocardial infarction (n, %) | 3 (6.8%) |
| Allergic (n, %) | 8 (18.2%) |
| Arrhythmology (n, %) | 30 (68.2%) |
| Diabetes mellitus (n, %) | 6 (13.64%) |
| Family history of autoimmune disease (n, %) | 0 |
| History of autoimmune disease (n, %) | 0 |
| Elevated myocardial enzyme profile (n, %) | 13 (29.5%) |
| NYHA | |
|    II | 2 (4.55 %) |
|    III | 14 (31.82 %) |
|    IV | 20 (45.45 %) |
| LVEDD | 63.60 ± 14.52 |
| LVEF | 32.00 ± 13.84 |

n, Patient number; NYHA, New York Heart Association; LVEDD, Left Ventricular End-Diastolic Diameter; LVEF, Left Ventricular Ejection Fraction.

**3.2 Performance of image analysis**

**Figure 3** illustrates an example that highlights the effectiveness of the StarDist network in the identification of nuclei. The patch selected for analysis was obtained from the region of myocardial inflammatory infiltration in patients with giant cell myocarditis. All nuclei with an accuracy of 0.982.

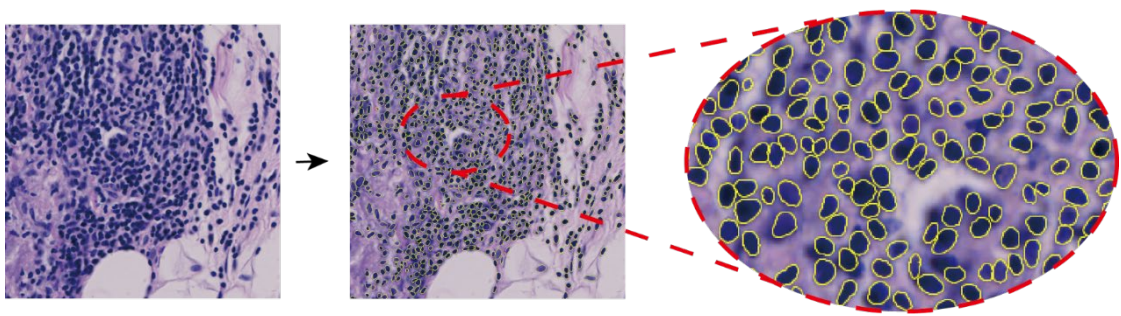

**Figure 3** Nuclear identification in the region of myocardial inflammatory infiltration.

**Figure 4** illustrates an example of the automatic identification of myocardial inflammatory infiltration. Myocardial nuclei and fiber nuclei were excluded by setting a radius from **Figure 4(a)** to **Figure 4(b)**. In **Figure 4(c)**, the area enclosed by the solid yellow line represents the identified myocardial inflammatory infiltration. **Table 3** shows the accuracy of our identification algorithm. It could accurately identify different types of myocarditis. The accuracy for the cases of DCM without myocarditis was above 0.99 and it is much higher compared to those cases with myocarditis. Among the various types of myocarditis, eosinophilic myocarditis exhibited the lowest accuracy (0.926), while DCM with focal lymphocytic myocarditis demonstrated the highest accuracy (0.997). The average accuracy for identifying myocardial inflammatory infiltration was 0.973.

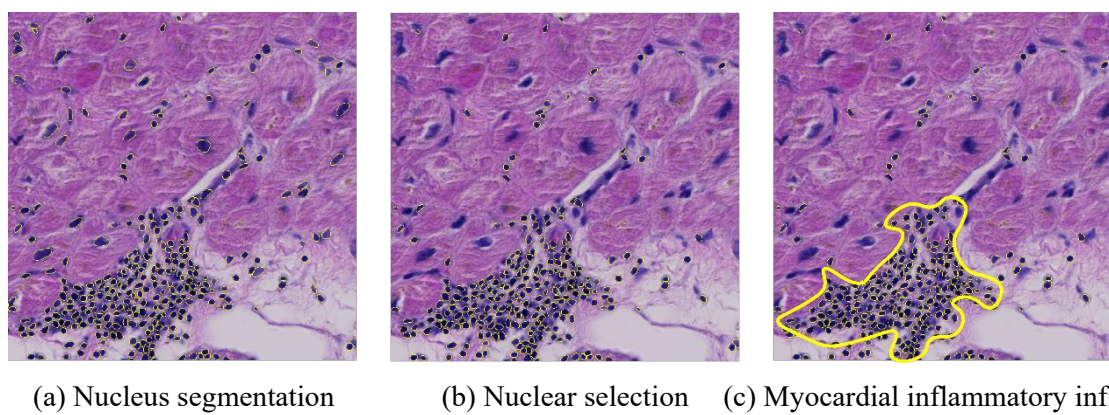

(a) Nucleus segmentation   (b) Nuclear selection   (c) Myocardial inflammatory infiltrate

**Figure 4** The process of myocardial inflammatory infiltration identification. (a) is the result of nuclear segmentation, (b) shows the result of simple nuclear selection, and (c) the area of the solid yellow line is the identified area of myocardial inflammatory infiltration.

**Table 3** Myocardial inflammatory infiltration identification results of the ten selected WSIs

| ID | Type | Accuracy |
|----|------|----------|
| 1 | DCM without myocarditis | 0.997 |
| 2 | DCM without myocarditis | 0.999 |
| 3 | DCM without myocarditis | 0.992 |
| 4 | Lymphocytic myocarditis | 0.939 |
| 5 | Eosinophilic myocarditis | 0.926 |
| 6 | Granulomatous myocarditis | 0.939 |
| 7 | Giant cell myocarditis | 0.957 |
| 8 | DCM with neutrophil myocarditis | 0.993 |
| 9 | DCM with focal lymphocytic myocarditis | 0.997 |
| 10 | Chronic active lymphocytic myocarditis | 0.988 |
| Total |  | **0.973** |

### 3.3 Diagnosis performance

By examining the diagnosis performance on the training set, it was observed that when the LND was above 1.7, all samples were diagnosed as positive, though the samples were mostly diagnosed as negative when the LND was below 0.6. Diagnoses were performed at intervals of 0.01 within the threshold range of 0.6 to 1.7. It was found that the diagnosis performed better at a cutoff value for LND of 1.02/mm$^2$.

Table 4 Performance evaluation results on the internal and external test sets

| Type | Five-fold cross-validation | Internal validation Set | External testing Set |
| --- | --- | --- | --- |
| AUC | 0.849±0.044 | 0.854 | 0.852 |
| Accuracy | 0.899±0.035 | 0.887 | 0.853 |
| Sensitivity | 0.971±0.017 | 0.971 | 0.845 |
| Specificity | 0.728±0.073 | 0.737 | 0.858 |

LND = 1.02/mm$^2$, data of five-fold cross-validation are expressed as mean ± Standard deviation.

The model performance of the five-fold cross-validation is depicted in **Table 4**. The results indicate that the standard deviation is relatively small. Both the accuracy and AUC values are over 0.84.

Based on **Table 4**,, the AUC for the internal and external test sets were 0.854 and 0.852, respectively, at the level of WSIs. The AUC performance remains stable for both the internal and external test sets. And the accuracy is consistent across the two test sets.

At the individual patient level, a positive diagnosis of myocarditis is determined by the presence of a positive biopsy result. In the myocarditis group, the automated methods correctly identified all patients as positive. In the double-blind trial group comprising 110 patients, there were 11 patients with false diagnostic results. Among these, 10 cases were false positives (0.091), and one was false negative (0.009).

## 4. Discussion

To our knowledge, this is the first study of DL-based computational pathology in the field of myocarditis. Our DL-based approach automatically identified nuclei and myocardial inflammatory infiltration and then measured LND in HE-stained images. The LND had high accuracy and robustness in the diagnosis of myocarditis.

**4.1 Computational pathology for the diagnosis of myocarditis**

Myocarditis can be diagnosed by either histological or immunohistochemical criteria[3,18]. The immunohistochemical criteria have not yet been incorporated into clinical guidelines[4]. Moreover, they require multiple antibody tests to determine the specific type of myocarditis[19], leading to experimental errors and increased medical costs. Histological criteria serve as the gold standard for the pathological diagnosis of myocarditis. HE-staining, a simple and widely used method, provides comprehensive histological information by staining various components of tissue cells. However, diagnosing myocarditis based on HE-stained images is a time-consuming and labor-intensive task, heavily reliant on the expertise and interpretation of pathologists, which can vary widely between observers. Therefore, there is an increasing need for an automated quantitative diagnosis of myocarditis using HE-stained images.

Computational pathology has significantly improved diagnostic performance and reduced human error rates in the automatic analysis of HE-stained images, including prostate cancer[20], colorectal cancer[21,22], and breast cancer[23]. ML, especially DL, has made remarkable progress in nuclei detection and classification[24,25] and tissue recognition[26,27] within the field of oncology. In the cardiovascular field, Nirschl et al.[28] developed a CNN classifier that outperformed pathologists in detecting clinical heart failure from cardiac histopathology, with 20% higher sensitivity and specificity. Peyster et al.[17] devised an ML-based grading method for cardiac allograft rejection, demonstrating agreement rates of 65.9% and 62.6% with the recorded grade and human graders, respectively. Lipkova et al.[29] developed a DL model for immune rejection detection and grading, achieving an AUC of 0.962 for allograft rejection detection and an AUC of 0.833 for distinguishing between low-grade and high-grade rejections. AI-based approaches, especially DL-based methods, have proven to be on par with traditional evaluation methods for HE-stained images, reducing inter-observer variability and evaluation time. These studies provide a solid foundation for the development of a computational pathology approach that enables automatic quantitative diagnosis of myocarditis using HE-stained images.

### 4.2 Our approach to the diagnosis of myocarditis

A DL-based computational pathology approach was developed to automatically identify and quantify myocardial inflammatory infiltration in digital HE-stained images in this study. Our approach utilizes the StarDist network interface for nuclear identification, which has shown excellent performance with an accuracy of 0.982. Building upon the nuclear detection results and leveraging the biological characteristics of myocardial inflammatory infiltration, our algorithm could automatically and accurately identify inflammatory infiltrates (accuracy: 0.973).

LND was quantified using this DL-based computational pathology approach. It was proven to be a valuable predictor for the diagnosis of myocarditis. The fivefold cross-validation analysis in **Table 4** confirmed that the diagnostic threshold of LND can be established by training with a limited dataset, resulting in high accuracy and stability. Moreover, the results from the internal and external test sets in **Table 4** emphasized the reliable and consistent diagnostic ability of LND in accurately differentiating myocarditis.

At the WSIs level, the sensitivity and specificity of the internal test group were significantly different (**Table 4**), while they were similar in the double-blinded group. This may be caused by the proportion of negatives and positives in the two groups and the small sample size of the internal test set.

In the double-blind trial group, the rate of false positives (0.091) was much higher than the rate of false negatives (0.009). This discrepancy is mainly due to the presence of perivascular tissue. A small number of myocarditis slices were misinterpreted as inflammatory infiltration. This error can be brought by the StarDist network segmentation algorithm. A technique, non-maximum suppression (NMS), is used in the StarDist segmentation algorithm to produce smoother nuclear shapes. The utilization of NMS can improve the accuracy of identifying the nuclei but also lead to inaccurate recognition of the edges of irregularly-shaped fiber nuclei surrounding blood vessels, thereby causing misidentifications. Nevertheless, visually reviewing the identified patches of myocardial inflammatory infiltrates can promptly resolve this error, resulting in a more accurate myocarditis diagnosis.

Our approach can be further improved by enhancing the nuclei segmentation. Note that the StarDist segmentation is a CNN-based model trained by pathology images from cancer patients. In our future work, we will apply transfer learning to the StarDist network and our myocardial pathology images to develop a more accurate segmentation model dedicated to myocardial cells.

### 4.3 Clinical use of our approach

Our approach has the potential to help pathologists quickly diagnose myocarditis based on the histopathology of heart tissue from biopsies, heart transplants, and autopsies. Because immune infiltration is diffuse, we need more evidence to make a negative diagnosis of myocarditis. In this condition, the pathologist should scrutinize every detail of the WSI, which introduces some false negative errors. These errors, in turn, cause more problems in clinical practice and judicial evaluation.

Besides, our approach offers the other two potential applications. One is the automatic quantification of LND from HE-stained images. Our study suggests that LND serves as an excellent predictor for diagnosing myocarditis. This application facilitates the objective and efficient assessment of LND, aiding in the accurate diagnosis of myocarditis. The other application is the automatic identification of myocardial inflammatory infiltration. Our algorithm allows clinics to swiftly identify and locate abnormal areas within the myocardium based on our identification results. This capability enhances diagnostic efficiency and reduces the likelihood of missed diagnoses. Besides, it can be used equally well for the detection and classification of myocardial immune rejection. By automating this process, our approach can save valuable time for clinicians and improve the overall diagnostic workflow.

### 4.4 Limitations

First, this study performed analysis on a relatively small number of patients (154). Second, our approach used samples from heart transplant patients rather than samples from EMB. The current gold standard for diagnosing myocarditis is commonly based on EMBs. However, due to the unavailability of EMBs, only samples from heart transplant patients were used. Future studies could investigate the feasibility and performance of our method on EMB samples to validate its effectiveness further and directly compare it with the current gold standard. Third, the approach was validated only with WSIs generated from Zeiss microscopes. Further studies with WSIs generated from other microscopes are warranted in future studies.

## 5. Conclusion

Our method can automatically identify and quantify myocardial inflammatory infiltration in HE-stained images. The new predictor LND is accurate and reliable in diagnosing myocarditis. Integrating our automated approach into clinical practice can potentially improve the efficiency and accuracy in the diagnosis of myocarditis.